\documentstyle[12pt]{article}
\def\lsim{\mathrel{\rlap{\raise 2.5pt \hbox{$<$}}\lower 2.5pt}}
\def\gsim{\mathrel{\rlap{\raise 2.5pt \hbox{$>$}}\lower 2.5pt}}
\begin{document}
\thispagestyle{empty}
\begin{small}
\begin{flushright}
UNIL-TP-1/95, NORDITA - 95/20 P, hep-ph/9503323\\
\end{flushright}
\end{small}
\vspace{5mm}
\begin{center}
{\bf{\Large The Non-Minimal Supersymmetric Standard
Model with $\tan\beta\simeq m_t/m_b$}}
\vskip 0.5cm

B. Ananthanarayan\\ [-2mm]
Institut de physique th\'eorique, Universit\'e de Lausanne,\\ [-2mm]
CH 1015, Lausanne, Switzerland.\\
\vskip 0.2cm

P. N. Pandita\\[-2mm]
NORDITA, Blegdamsvej 17, DK-2100, Copenhagen {\O}, Denmark\\[-2mm]
and\\[-2mm]
Department of Physics, North Eastern Hill University,\\[-2mm]
Laitumkhrah, Shillong 793 003, India.$^*$
\end{center}
\begin{abstract}
We consider the supersymmetric extension of the standard
model with an additional singlet $S$, the Non-Minimal Supersymmetric
Standard Model (NMSSM),
in the limit $\tan\beta \simeq m_t/m_b$.
We embed this model in a supergravity framework with
universal boundary conditions and analyze the renormalization
group improved tree-level potential.
We examine the relationship between this model and
the minimial supersymmetric standard model (MSSM),
and discuss the novel connections between the two when $\tan\beta$
is large.
Strong correlations
between the free parameters of the nonminimal model are found and
the reasons for these discussed.
The singlet vacuum expectation value is forced to be large,
of the order of $10\ TeV$.  The radiatively corrected
mass of the lightest Higgs boson is found
to be $\stackrel{_<}{_\sim} 140\ GeV$.
\end{abstract}

\noindent{\underline{\hspace{11.6cm}}}\\
\begin{small}
* Permanent address
\end{small}


\newpage
\noindent{\bf 1. Introduction.}

\bigskip

The recent past has witnessed much activity in exploring
supersymmetric unification[1].   Furthermore, improvements
in the determinations of the standard model couplings
have given us reason to believe that supersymmetric
unification with a SUSY breaking scale of
$\sim 1\ TeV$ is  compatible with these
measurements[2]. Other predictions from
supersymmetric unification are dogged by the lack of
knowledge of the crucial parameter $\tan\beta\equiv
v_2/v_1$, the ratio of the vacuum expectation values of
the two Higgs doublets $H_2$ and $H_1$ required to give
masses to the up-type and the down-type (and charged
leptons) quarks, respectively (our normalization is
$\sqrt{v_1^2+v_2^2}=174 GeV$ and the mass of the Z boson
is defined such that
$m_Z^2=\frac{1}{2}(g^2+g'^2)v^2$, where $g$ and $g'$ are
the gauge couplings of $SU(2)$ and $U(1)$, respectively).

One particularly predictive framework
is based on the assumption that the heaviest generation
fermions lie in a unique {\bf 16}-dimensional representation
of the unifying gauge group $SO(10)$ with the Higgs doublets
in a {\bf 10}-dimensional representation of the group[3].  This
implies that the top-quark, b-quark and $\tau$ lepton
Yukawa interactions arise from a $h. {\bf 16}.{\bf 16}.{\bf 10}$ term in
the superpotential at the unification scale $M_X$ determined
from gauge coupling unification.
The coupled system of differential equations for
the gauge couplings and Yukawa couplings are then
evolved down to present energies from $M_X$,
and $\tan\beta$ is determined from the accurately
measured value of $m_\tau=1.78\ GeV$.  When $h(=h_t=
h_b=h_\tau)$ is chosen in such a manner as to yield
a value for $m_b(m_b)$ in its ``observed'' range of
$4.25\pm 0.01\ GeV$[4], a rather good prediction for the
top-quark mass parameter $m_t(m_t)$ is obtained, which
with the present central value of $\alpha_S(M_Z)=0.12$
lies in the range favoured by the experimental data[5].
Here $\tan\beta$ is found to
saturate what is considered to be a theoretical upper bound on
its value of $m_t/m_b$ and the Yukawa coupling $h$ is found to come
out to be
rather large $O(1-3)$ with a certain insensitivity
to the exact value since it is near a fixed point of its
evolution.

In $SU(5)$ type unification
where $\tan\beta$ is free, the region $\tan\beta \approx 1$
is also a region which is favoured for the unification of
the b-quark and $\tau$-lepton masses from the observed data[6].
One crucial difference between the two extremes discussed
above is that in the $SO(10)$ case the Yukawa couplings of
the b-quark (and that of the $\tau$ lepton) always remain
comparable to that of the top-quark, with the observed
hierachy in the masses of these quarks arising from the
large value of $\tan\beta$, while in the $SU(5)$ case
the Yukawa couplings of the b-quark and the $\tau$-lepton
are negligible in comparison with that of the top-quark.

The above discussion about unification does not involve in any great detail
the remaining aspects of the embedding of the standard model
into a supersymmetric grand unified framework.  The minimal
supersymmetric extension
of the standard model requires, besides the superpartners, the introduction
of an additional Higgs
doublet,
and indeed with this matter content and an additional
symmetry known as matter parity [1], to forbid couplings
that lead to rapid nucleon decay, it is possible to construct
a self-consistent and highly successful
framework which has come to be known as
the Minimal Supersymmetric Standard Model (MSSM)[1].  Here the mass
of the lightest  Higgs boson, which is unknown in the Standard Model,
is related,
through the D-term in the potential, to the mass of the
Z boson, $m_Z$. It is also related to $\tan\beta$ and $m_A$
($>>m_z$ as required by $b\rightarrow s\gamma$ constraints),
the mass of the CP-odd neutral scalar boson that remains as a physical
degree of freedom
after the breakdown of $SU(2)\times U(1)$.  It has been shown
that the mass of the lightest (``Weinberg-Salam'') Higgs
$m_{h^0}$ in the MSSM for  large values of $\tan\beta$, after
inclusion of
radiative corrections due to the presence of large top Yukawa coupling,
is $\stackrel{_<}{_\sim} 140\ GeV$[7].
Furthermore, supersymmetry breaking is understood to
arise from embedding MSSM into a supergravity framework
and writing down all the possible soft-supersymmetry breaking
terms consistent with the gauge and
discrete symmetries that define the model.  It is often assumed
that many of the parameters describing these terms are in fact
equal at the unification scale in order to have a predictive framework.
Such universal conditions have had to be relaxed
in order to avoid fine tuning and minimize the
effects of possible radiative corrections to the b-quark
mass due to the fine details of the spectrum of this model,
especially in the case when $\tan\beta$ is large[7,8].

Despite its many successes it may be premature
to confine our attention only to the MSSM, especially because of the
presence of the dimensionful Higgs bilinear parameter $\mu$ in the
superpotential.
An alternative model to MSSM that has widely been considered is the  one
where the
Higgs content is  extended (economically) by the addition
of a gauge singlet $S$, and assuming a discrete
$Z_3$ symmetry in order to avoid
linear and bi-linear couplings in the superpotential[9], the so called
Non-Minimal Supersymmetric Standard Model(NMSSM).
This model is also referred to as the next to minimal supersymmetric
model or as (M+1)SSM.  In particular, this corresponds to
forbiding the $\mu H_1.H_2$ coupling of the MSSM in the
superpotential and instead introduce the couplings
\begin{equation}
\lambda S H_1.H_2+\frac{1}{3}k S^3
\end{equation}
 with the effective
``$\mu$'' term generated by the vacuum expectation value
$<S>(\equiv s)\neq 0$.  This model is particularly interesting since
it does not affect the positive features of the MSSM including
gauge coupling unification [10] and allows a
test of the stability of the features of the MSSM such
as the upper bound on the mass of the lightest Higgs boson
with favourable results.
It has a significantly richer phenomenology and a typically
larger parameter space[11,12,13].

However, in recent studies of this model with
high energy inputs[11,14],
the interesting case of large $\tan\beta$ has not been considered.
In this paper we discuss this possibility: that of the non-minimal
supersymmetric standard model with
large $\tan\beta$. We note that there is a distinct possibility that
the desirable features of a good prediction for the top-quark
mass may in fact be destroyed by the presence of additional
Yukawa couplings $\lambda$ and $k$
which couple to the Yukawa couplings of the
heaviest generation even at one-loop level, but in actual practice has
been found not to be the case[15].

 We shall discuss different features
of the non-minimal supersymmetric standard model at large values
of $\tan\beta$ ,
and compare and contrast these, as often as possible,
with the corresponding features in MSSM. We shall also discuss
under what conditions one can
obtain the latter as a well-defined limit of the former.
We have carried out a renormalization group analysis of
this model with universal boundary conditions and analyzed the
renormalization group improved tree-level potential at the
scale $Q_0$. The cut off scale for the renormalization group evolution
is chosen to be the geometric mean of the scalar top quark masses which
is roughly equal to that of the geometric mean of the scalar b-quark
masses as well, since during the course of their evolution the Yukawa
couplings of the t and b-quarks are equal upto their hypercharges and the
relatively minor contribution of the $\tau$-lepton.
Whereas in the MSSM the parameters $\mu$ and $B$
(the soft susy parameter characterizing the bilinear term in
the scalar potential) do not enter into the evolution of the
other  parameters of the model at one-loop level, the situation
encountered here is drastically different with a systematic
search in the parameter space having to be performed with
all parameters coupled from the outset.
Our analysis of the minimization conditions
that ensure a vacuum gives rise to severe fine tuning problems,
that persist in this model, as in the MSSM.
The problems are further compounded by having the constraints of
three minimization conditions, rather than two such conditions
that occur in MSSM.  In previous studies of the model where $\tan\beta$ was
free,  the tuning of parameters
was possible in order to meet all the requisite criteria, viz.,
minimization conditions, requirement that the vacuum
preserve electric charge and colour, etc.  However, in
the present where $\tan\beta$ is fixed and large, what
we find is
a highly correlated system. An important conclusion that we draw here
is that just as in the case of MSSM, for large values of $\tan\beta$,
the upper bound on the mass of the
lightest Higgs boson
is $\stackrel{_<}{_\sim}140\ GeV$.

\bigskip

\noindent{\bf 2. The Model}

\bigskip

The model is characterized by the following couplings in the
superpotential, where we exhibit the  interactions of the
heaviest generation and
the Higgs (singlet and doublet) sector of the theory:

\begin{equation}
W=h_t Q\cdot H_2 t^c_R+h_b Q\cdot H_1b^c_R+h_\tau L\cdot H_1\tau^c_R+
\lambda S H_1.H_2+\frac{1}{3}kS^3
\end{equation}

\noindent One has to add to the potential obtained from (2) the most
general terms
that break supersymmetry softly.
We explicitly
write down these soft breaking terms here, since it will establish our sign
conventions for the relevant parameters. The potential can be computed
from (2) by a standard procedure[11]. The relevant soft susy breaking
terms are:
$$
(h_tA_t \tilde{Q}\cdot H_2 \tilde{t^c_R}+
h_b \tilde{Q}. H_1 \tilde{b^c_R} +
h_\tau A_\tau \tilde{L} \cdot H_1\tilde{\tau^c_R} +
\lambda A_\lambda H_1\cdot H_2 S +
{{1}\over{3}}k A_k S^3)+ {\rm h. c.}
$$
$$
+m_{H_1}^2|H_1|^2+m_{H_2}^2|H_2|^2+m_S^2|S|^2
+m_{\tilde{Q}}^2|{\tilde{Q}}|^2+
m_{\tilde{t}}^2|{\tilde{t}}^c_R|^2
+m_{\tilde{b}}^2|{\tilde{b}}^c_R|^2
+m_{\tilde{\tau}}^2|{\tilde{\tau}}^c_R|^2.
$$

The conventions for the gaugino masses follow those of the
MSSM[16].  Since part of the discussion that follows rests
on the minimization conditions
(evaluated at $Q_0$ after all the parameters are evolved
via their one-loop renormalization group equations
down to this scale) we give them here:
\begin{eqnarray}
m_{H_1}^2=-\lambda{{v_2}\over{v_1}} s (A_\lambda+ks) -\lambda^2(v_2^2+s^2)
+{{1}\over{4}}(g^2+g'^2)(v_2^2-v_1^2) \\
m_{H_2}^2=-\lambda {{v_1}\over{v_2}} s (A_\lambda+ks)-\lambda^2(v_1^2+s^2)
+{{1}\over{4}}(g^2+g'^2)(v_1^2-v_2^2)\\
m_{S}^2=-\lambda^2(v_1^2+v_2^2)-2k^2s^2-2\lambda s v_1 v_2 - kA_k s -
{{\lambda A_\lambda v_1 v_2}\over {s}}
\end{eqnarray}

\noindent One may rewrite the first two minimization equations to obtain:
\begin{eqnarray}
\tan^2\beta={{m_Z^2/2+m_{H_1}^2+\lambda^2s^2}\over
{m_Z^2/2+m_{H_2}^2+\lambda^2s^2}}  \\
\sin 2\beta={{(-2\lambda s)(A_\lambda+k s)}\over{m_{H_1}^2
+m_{H_2}^2+\lambda^2(2s^2+v^2)}}
\end{eqnarray}

Equations (6) and (7) give us some crucial
insights into the manner in which our solutions may behave like.  The
first of these guarantees that, as in the MSSM, $\tan\beta$
must lie between 1 and $m_t/m_b$.  The proof of this relies
once more on the renormalization group equations that
govern the behaviour of the mass parameters and may be
proved very simply by reductio ad absurdum.  For this
purpose we need only consider the following equation expressing
the momentum dependence of the  difference of two supersymmetry
breaking scalar mass  parameters:
\begin{equation}
{{d}\over{dt}}(m_{H_1}^2-m_{H_2}^2)={{1}\over{8\pi^2}}
(-3h_t^2X_t+3h_b^2X_b+h_\tau^2X_\tau)
\end{equation}
\noindent where $t=\log(\mu)$, the logarithm of the
momentum scale, and $X_i$, $i=t,\ b,\ \tau$, are
combinations of scalar masses and tri-linear couplings:
\begin{eqnarray}
X_t=m_{\tilde{Q}}^2+m_{\tilde{t}}^2+m_{H_2}^2+A_t^2 \nonumber \\
X_b=m_{\tilde{Q}}^2+m_{\tilde{b}}^2+m_{H_1}^2+A_b^2 \nonumber \\
X_\tau=m_{\tilde{L}}^2+m_{\tilde{\tau}}^2+m_{H_1}^2+A_\tau^2 \nonumber
\end{eqnarray}
It must be
noted that in order to prove that $\tan\beta > 1$ we neglect
$h_b$ and $h_\tau$ and for proving $\tan\beta < m_t/m_b$
we retain them.

The $\tan^2\beta$ equation (6) shows that, as in the MSSM, in
order to guarantee a large $\tan\beta$, with the
essential degeneracy of $m_{H_1}^2$ and $m_{H_2}^2$ enforced
by the renormalization group equations, the denominator of
the equation has to come out to be small at low scale.  Therefore,
we have the fine-tuning condition that
\begin{equation}
m_{H_2}^2+\lambda^2s^2 \approx -m_Z^2/2
\end{equation}

\noindent Here one notes that the correspondence with the MSSM will occur
in a certain well defined manner with the identification of
$\lambda s$ with $\mu$.  Similarly one has to identify
$A_\lambda+ks$ with $B$.  What we will show below is that in
the large $\tan\beta$ case this identification occurs in a
novel way that is not generic to the model, say, in the limit
of $\tan\beta \approx 1$.

An investigation of the $\sin 2\beta$  equation yields, when
$\beta\approx \frac{\pi}{2}$, as in the case at hand, that one
must have the condition
\begin{equation}
A_\lambda\approx -ks
\end{equation}
This is similar to the condition in the MSSM that $B\approx
0$.  Here the situation is far worse since $A_\lambda$ is not
a parameter that is fixed at $Q_0$ but is present from
the outset.  This is the first of the fine tuning problems
that we encounter.

A rearrangement of the $\sin 2\beta$ equation yields:
\begin{equation}
\lambda s (A_\lambda+ks)=\tan\beta
(-m_{H_2}^2-\lambda^2s^2){{m_Z^2}\over{2}}{{(\tan\beta^2-1)}
\over{(\tan\beta^2+1)}}-{{\lambda^2v^2\sin 2\beta}
\over{2}}
\end{equation}

In this equation it is legitimate to discard the last term
for the case of large $\tan\beta$ and one sees here that with the
 identification
of the appropriate parameters in terms of the MSSM parameters as
described earlier, one recovers all the analogous MSSM relations
for all values of the other parameters without having to go
through a limiting procedure[11], as is the case when $\tan \beta$
is arbitrary.

The next fine tuning condition we encounter is related to
the third minimization condition which we rewrite as:
\begin{equation}
m_S^2=-\lambda^2 v^2-2 A_\lambda^2+{{\lambda v^2 \sin 2\beta A_\lambda}
\over{s}}+A_\lambda A_k +{{\lambda k v^2 \sin 2\beta}\over{2}}
\end{equation}

Here one may observe that in order to satisfy this condition
one must have large cancellations between the fourth and the
first two terms since the terms proportional to $\sin 2\beta$
are negligible.  This requires that $A_k$ and $A_\lambda$
come out with the same sign and that the product be sufficiently
large.  As we shall see, it is this condition that leads to
problems with finding solutions with sufficiently small tri-linear
couplings in magnitude.

\bigskip

\noindent{\bf 3. Results and Conclusions}

\bigskip

The starting point of the program is the estimation of the scale
$M_X$ with the choice of the SUSY breaking scale $Q_0\sim 1\ TeV$.
For $\alpha_S(m_Z)=0.12$, $Q_0 = 1\ TeV$ and $\alpha=1/128$,
we find upon integrating the one-loop beta functions, $M_X=1.9\times
10^{16}\ GeV$ and the unified gauge coupling $\alpha_G(M_X)=1/25.6$.
We then choose a value for the unified Yukawa coupling $h$ of $O(1)$.
The free parameters of the model are $(M_{1/2},\ m_0,\ A,\ \lambda,\ k)$,
which are the common gaugino mass, the common scalar mass, the
common tri-linear scalar coupling and the two additional Yukawa couplings
respectively.
Note that our convention requires us to choose $\lambda>0$ and
$k<0$ in order to conserve CP in the Yukawa sector of the model[11].

We then write down the coupled system of renormalization group (RG)
equations for the 24 parameters of the model that are coupled to each other
(ignoring the parameters of the lighter two generations since they
do not couple to the rest of the parameters at the one-loop level)
and evolve this system down to present energies of $Q_0$.
These RG equations may  be obtained by generalizing the
expressions of
Ref. [9] to include the contributions of
$h_b$ and $h_\tau$ and  can derived from the general expressions of
Ref. [17].
We then compare the numerical values of the mass parameters that
enter the left
hand side of the minimization equations (3) - (5) with
the combination of the
parameters that enter the right hand side of these  equations as
obtained from the RG evolution.

In practice, it turns out that the first of the minimization
conditions, eq.(3), is the
most sensitive to the choice of initial conditions.   This
reflects the fine tuning condition that we dwelt on in the
previous section.   We also impose the
constraint that $|A|<3m_0$[18] in order to guarantee the absence of
electric-charge
breaking vacua.  In the case at hand this choice may have to
be strengthened further due to the presence of large Yukawa couplings
for the b-quark.  The situation is considerably less restrictive
when mild non-universality is allowed and, for instance, if strict Yukawa
unification is relaxed.  Given these uncertainties, we choose to work
with this constraint.

In Table 1 we present various sets of values of the input parameters
that we take and the corresponding output.  Our choices
are arranged in such a manner so  as to show the change in some of the
crucial features as we vary certain input parameters, some of which
are changed as we go down the columns.  Besides the input values
we present in Table 1 (a)
 certain crucial quantities that are calculated from the
renormalization group evolution at $Q_0$.  These are $\tan \beta$,
$r(\equiv s/v)$, $A_\lambda$, $A_k$ and $ks$.  In Table 1 (b), we
present the values of $r_1$, $r_2$ and $r_3$ which are defined as
the difference between the left and right hand sides of the
three minimization equations (3), (4) and (5) divided by the
right hand side of
each of these equations.  A genuine vacuum corresponds to all three being
equal to 0.  However, given the extremely fine tuned nature of these
conditions, we present points in the parameter space where their
variations are most easy to observe.  The choices of interest are
precisely those where these suffer a change in sign as one of the
parameters is changed indicating that such points lie in the
neighbourhood of the desired vacuum.
In Table 1(b) we  then present the corresponding
values of $\Delta E$, the difference in the value
of the scalar potential computed with the scalar fields attaining
their vacuum expectation values ($v_1$, $v_2$ and $s$) and its value
computed at the origin.
A negative $\Delta E$ signifies that the $SU(2)\times
U(1)$ breaking vacuum
has lower potential energy than the symmetric state $v_1=v_2=s=0$
(since this state is normalized to have a vanishing potential energy),
and is favoured.
 We also compute the mass of the charged Higgs boson[11]
\begin{equation}
m_C^2=m_W^2-\lambda^2v^2-\lambda(A_\lambda+ks){{2s}\over{\sin 2\beta}}
\end{equation}
\noindent where $m_W$ is the mass of the W-boson. We note that the
radiative corrections to the charged Higgs mass are small for most of
the parameter range, as in the case of MSSM[19]. The reason for this
is that a global $SU(2)\times SU(2)$ symmetry [20] protects the charged
Higgs mass from obtaining large radiative corrections.  If this quantity
were to come out to be negative, then the resulting vacuum would
break electric-charge spontaneously and the corresponding point
in the parameter
space would be excluded.  We also compute the squared masses
of the neutral pseudoscalar
bosons which are assured to be positive[11] for a
true vacuum.  However, we find that due to the fine-tuning
conditions,  as we move  away from the region where
a vacuum is to be found, the smaller eigenvalue of the mass squared
matrix of pseudoscalar Higgs bosons in fact changes sign.

We finally present in the last column of the table the quantity
that is of the greatest importance, viz., the upper bound (denoted by
$M_{h^0}$) on the
mass of the lightest Higgs boson in the model,
$m_{h^0}$. Including radiative corrections, this upper
bound can be written as[21]:
\begin{eqnarray}
\displaystyle m_{h^0}^2
& \stackrel{_<}{_\sim} &
 m_Z^2(\cos^2 2\beta+{{2\lambda^2}\over{g^2}}\sin^2 2\beta) \nonumber \\
& & \displaystyle +{{3g^2}\over{16\pi^2m_W^2}}(\Delta_{11} \cos^2\beta+
\Delta_{22}\sin^2\beta+\Delta_{12}\sin 2\beta)
\end{eqnarray}
\noindent where the $\Delta_{ij},\ i,j=1,2$ are the terms
that arise from the one-loop radiative corrections and involve
$m_t^4/m_W^2$, $m_b^4/m_W^2$ and logarithms of the squark masses.
For each of our inputs, we compute
the masses of the physical squark eigenstates in order to estimate
the upper bound (14).
We note that
in the limit $\beta \approx {{\pi}
\over{2}}$, the second term in the first bracket, which is proportional to
$\sin^2 2\beta$, is small, so that the upper bound (14) reduces to
the corresponding bound on the lightest Higgs mass in the MSSM when
appropriate identification of parameters is made. We further note that
the bound (14) depends only logarithmically on $ r$, and hence on
the singlet vacuum expectation value $s$, in the limit of large $r$,
which, therefore, decouples from the bound [22].

We broadly discuss the solutions that we find by first noting that the
solutions are easier found with $\lambda$ and $k$ chosen to be
smaller than 1.  Larger values increase their importance in the
renormalization group equations of the soft parameters.  There
is a rough scale invariance enjoyed by the solutions  when each of
these is scaled by the same parameter.  The same is also true of
scaling $M_{1/2},\ m_0$ and $A$.  Furthermore if $M_{1/2}$ is taken
to be negative and of the same order as $m_0$ and $A=3 m_0$,  there is
only a weak dependence on the actual value of the parameter.
Note that in  MSSM $m_0$ is required to be smaller than $M_{1/2}$
due to the correlations between these parameters and the mass of
the only pseudoscalar Higgs boson in the spectrum.

For the first 11 rows in Table 1(a)
we have taken a unified Yukawa coupling $h=1.5$ which is a typical value
for this parameter that yields a successful prediction for the top-quark
mass[3,8] (when $m_b$ is evolved using two-loop
QCD evolution equations this yields $m_b(m_b)=4.09\
GeV$ and $m_t(m_t)=181\ GeV$, ignoring small
corrections due to the presence of the additional
Yukawa couplings $\lambda$ and $k$[15]).
This is changed to a somewhat smaller value of $1.0$ for
the next 4 rows (corresponding to $m_b(m_b)=4.28\ GeV$ and
$m_t(m_t)=177\ GeV$) and the last four rows to
$h=2.0$ (corresponding to $m_b(m_b)=4.00\ GeV$ and $m_t(m_t)=183\ GeV$).
The value of $\lambda$ alone is changed as we move
down from the 1st to the 3rd rows.  The desired features that
the $r_i\ (i=1,2,3)$ suffer change of signs are seen.  Note that
$\Delta E$ and the $m_C^2$
also change sign.  This represents the presence of an instability
in the vacuum and with the present accuracy it is not possible to
ascertain whether any genuine vacuum can be found in this neighbourhood.
(A related  observation that we make is that
in this region the mass squared of the lighter pseudoscalar neutral
boson is also found to suffer a change of sign, from positive to
negative, as the parameter
$\lambda$ is increased, as  would be expected due to the
fine tuning.)
The rows 4 - 6 are similar to 1 - 3 with a smaller value of $|M_{1/2}|$.
The qualitative features described for the first three rows persist
here and illustrate the weak dependence on the actual
value of this parameter.

 In rows 7 and 8 we show the
changes that occur as $\lambda$ is increased with a  fixed
ratio $|A/m_0|=2$, which is a factor
of $2/3$ smaller than for the first 6 rows.  It is seen that
$r_3$ never approaches the neighbourhood of $0$ although $r_1$ does
suffer a sign change.  This illustrates the need for a large
value for this ratio of nearly 3.  This feature is also observed
in the case of small $\tan\beta$[14].  In rows 9 and 10, we scale down
the values of the dimensional parameters compared to, say, rows 2 and 3
and varying $\lambda$ in a range that overlaps that of these rows.
Similar qualitative features are found to persist showing the rough
scale invariance of the system described earlier.  Similarly in row 11,
we scale down $\lambda$ and $|k|$ simultaneously by a factor
of 10 compared to row 9 so as  to demonstrate the relative independence
from the absolute values of these parameters.  Note that the computed
value of $r$ scales inversely by roughly the same factor.

The
next four rows 12-15 show the behaviour of our solutions for a somewhat
smaller value of $h$ and, qualitatively, the  features are no different.
The possible solutions now require a smaller value of the ratio
$\lambda/|k|$.  And, finally the last four rows 16-19 show
the behaviour for the largest
$h$ that we have chosen for the purposes of illustration.

A point that deserves emphasis is the fact that the quantity $r$
persistenly remains rather large corresponding to the vacuum
expectation value of the singlet, $s$,
being about $10\ TeV$.  This is substantially different from the
case of $\tan\beta \approx 1$[11]. This implies that for large $\tan \beta$,
the vacuum expectation value of the Higgs singlet is forced
to be rather large. We note that the singlet vauum expectation values
are not constrained by the experimental data.

As a  final, and the most significant, point we go to
the last column of Table 1(b) wherein the absolute upper bound
$M_{h^0}$ on the lightest Higgs mass is presented for the choice
of input parameters of Table 1(a).
In Fig. 1 we plot, for typical and reasonable
values of the input parameters in the region
where the vacuum is expected to lie, the upper bound on
the mass of the lightest Higgs boson as a function of the top quark
mass $m_t(m_t)$ in the range that is most favoured under
these boundary conditions[3,7,8,15].  Such a linear behaviour
has also been observed in the case of the MSSM[23].
It is seen that the upper bound never exceeds
$140\ GeV$.  We, therefore,
conclude that
as in the MSSM, even with values of $h$ close
to the perturbative bound of $3.3$, this bound is not likely
to exceed $140 \ GeV$.  It must be noted that in MSSM due to
the presence of
of fewer degrees of freedom at the outset and only
two  minimization conditions and only one  pseudoscalar Higgs
boson in the spectrum[7], the
constraining of the mass of the lightest higgs scalar is
relatively easier.

{}From the detailed discussion presented above, we see that the
non-minimal supersymmetric standard model rests on  a rather delicately
hinged system of equations and constraints.
It has been argued, and found true[23, 24], that the
minimization of the tree-level potential in the MSSM is consistent
with minimizing the one-loop corrected potential to within $\sim
20\%$ when the cutoff is chosen in the range of the geometric
mean of the scalar top-quark masses[18], and the situation in
the nonminimal model is expected to be no different.
Whereas the non-minimal model provides a
good testing ground for the stability of predictions of the MSSM,
in practice it deserves great care in its treatment.
For instance, whereas in the MSSM the mass of the only
pseudoscalar Higgs boson plays an important role in constraining
the allowed regions of the parameter space[7], the situation here
is significantly more complicated and no simple constraint emerges
from our analysis. It appears to us, through the above careful analysis
of stability and fine tuning in the nonminimal model, that the
framework may have to be broadened
via non-universality of soft breaking terms as has been done
recently in the case of MSSM.
It is hoped that the
present work might serve as a springboard for such
investigations.

\bigskip

\noindent{\bf Acknowledgements}:
PNP thanks NORDITA for hospitality during the course of this work.
BA thanks the Swiss National Science Foundation for support during
the course of this work. The work of PNP was supported by the
Department of Science and
Technology, Government of India.  We thank Q. Shafi for discussions,
D. Toublan and D. Rickebusch for
assistance with the manuscript.

\newpage

\newpage

\noindent{\bf Table Caption}

\bigskip

\noindent Table 1 (a). Typical values for inputs $M_{1/2},\ m_0,\
A,\ h, \ \lambda, \ k$ at $M_X$ and the corresponding low energy values of
$\tan\beta$, $r$, $A_\lambda$, $A_k$ and $ks$ (all masses in
units of $GeV$.).

\bigskip

\noindent Table 1 (b).  The values of $r_1,\ r_2,\
r_3, \ \Delta E$ and $m_C^2$ and the upper limit on
the mass of the lightest higgs $M_{h^0}$
for the inputs of Table 1a (all masses in
units of $GeV$.)

\bigskip
\bigskip

\noindent{\bf Figure Caption}

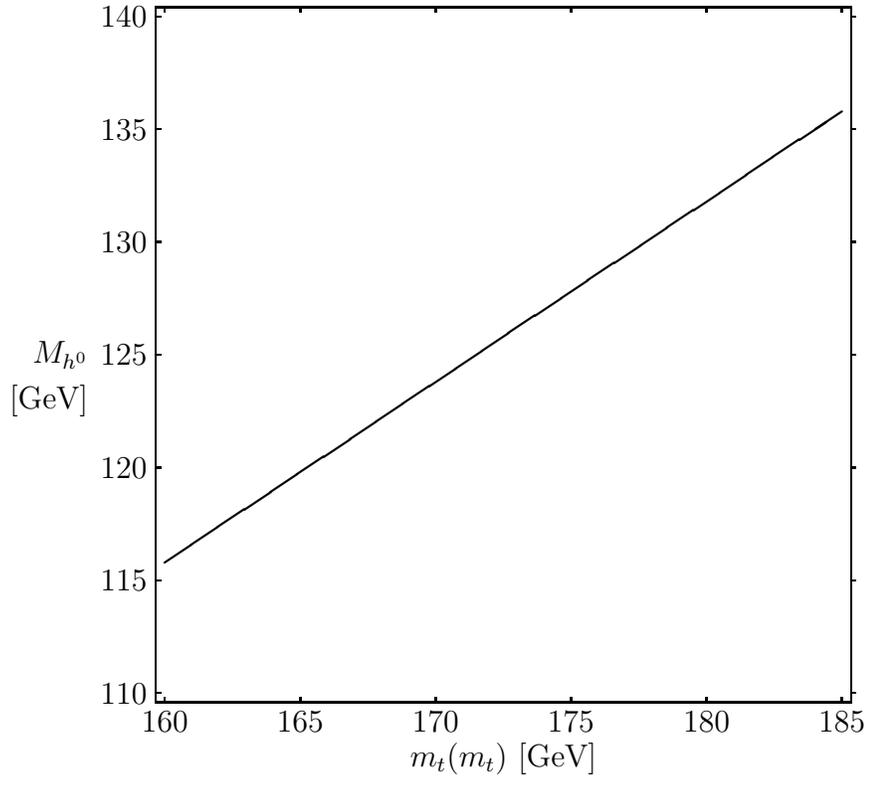
\noindent Fig. 1. Plot of upper bound on the mass of the lightest
higgs, $M_{h^0}$ vs. $m_t(m_t)$ for a typical choice of parameters,
$M_{1/2}=-700\ GeV,\ m_0=800\ GeV, A=2400\ GeV$ and $\lambda/|k|$
chosen to yield the neighbourhood of a vacuum.

\bigskip
\newpage

\begin{footnotesize}
$$
\begin{array}{||c|c|c|c|c|c|c||c|c|c|c|c||}\hline
\# & M_{1/2} & m_0 &  A & h & \lambda & k & \tan\beta & r & A_\lambda & A_k
& ks \\ \hline
1 & -700 & 800 & 2400 & 1.5 & 0.35 & -0.10 & 62 & 50 & 732 & 2257 & -828 \\
2 & -700 & 800 & 2400 & 1.5 & 0.40 & -0.10 & 62 & 44 & 718 & 2231 & -721 \\
3 & -700 & 800 & 2400 & 1.5 & 0.45 & -0.10 & 62 & 40 & 703 & 2203 & -637 \\
4 & -500 & 800 & 2400 & 1.5 & 0.35 & -0.10 & 62 & 40 & 577 & 2258 & -658 \\
5 & -500 & 800 & 2400 & 1.5 & 0.40 & -0.10 & 62 & 35 & 563 & 2233 & -573 \\
6 & -500 & 800 & 2400 & 1.5 & 0.45 & -0.10 & 62 & 32 & 548 & 2205 & -507 \\
7 & -700 & 800 & 1600 & 1.5 & 0.40 & -0.10 & 62 & 44 & 659 & 1485 & -708 \\
8 & -700 & 800 & 1600 & 1.5 & 0.50 & -0.10 & 62 & 36 & 637 & 1445 & -560 \\
9 & -350 & 400 & 1200 & 1.5 & 0.40 & -0.10 & 62 & 22 & 359 & 1115 & -358 \\
10 & -350 & 400 & 1200 & 1.5 & 0.50 & -0.10 & 62 & 18 & 343 & 1086 & -283 \\
11 & -350 & 400 & 1200 & 1.5 & 0.04 & -0.01 & 62 & 212 & 399 & 1199 & -368 \\
12 & -700 & 800 & 2400 & 1.0 & 0.20 & -0.10 & 58 & 59 & 835 & 2291 & -986 \\
13 & -700 & 800 & 2400 & 1.0 & 0.30 & -0.10 & 58 & 40 & 799 & 2226 & -652 \\
14 & -700 & 800 & 1600 & 1.0 & 0.20 & -0.10 & 58 & 57 & 722 & 1526 & -952 \\
15 & -700 & 800 & 1600 & 1.0 & 0.30 & -0.10 & 58 & 39 & 697 & 1483 & -629 \\
16 & -700 & 800 & 2400 & 2.0 & 0.50 & -0.10 & 64 & 47 & 685 & 2235 & -757 \\
17 & -700 & 800 & 2400 & 2.0 & 0.60 & -0.10 & 64 & 40 & 661 & 2191 & -624 \\
18 & -700 & 800 & 1600 & 2.0 & 0.50 & -0.10 & 64 & 46 & 643 & 1487 & -749 \\
19 & -700 & 800 & 1600 & 2.0 & 0.60 & -0.10 & 64 & 39 & 626 & 1457 & -617 \\
\hline
\end{array}
$$
\end{footnotesize}
\bigskip
$$
{\rm Table\ 1\ (a)}
$$

\begin{footnotesize}
\bigskip
$$
\begin{array}{|c|c|c|c|c|c|c||}\hline
\# & r_1 & r_2 & r_3 & \Delta E (\cdot 10^{11}) & m_C^2(\cdot 10^7) & M_{h^0}
\\ \hline
1 & 5.6 & 1.6\cdot 10^{-3} & 0.1 & -5.5 & 0.59 & 132 \\
2 & 0.3 & 7.8\cdot 10^{-5} & -4.2\cdot 10^{-2} & -4.2 & 0.02 & 132 \\
3 & -3.6 & -1.0\cdot 10^{-3} & -0.1 & 3.4 & -0.39  & 132 \\
4 & 5.5 & 1.7\cdot 10^{-3} & -0.1 & -2.4 & 0.39 & 128 \\
5 & 0.8 & 2.6\cdot 10^{-4} & -0.1 & 7.4 & 0.05 & 128 \\
6 & -2.5 & -7.7\cdot 10^{-4} & -0.1 & 14.3 & -0.20 & 128\\
7 & 3.0 & 8.3\cdot10^{-4} & 0.9 & 216.0 & 0.29 & 132 \\
8 & -4.4 & -1.3\cdot 10^{-3} & 0.7 & 128.0 & -0.46 & 132 \\
9 & 3.2\cdot 10^{-4} & 9.3\cdot 10^{-6} & -4.5\cdot 10^{-2} & -0.25 & -0.0016 &
122\\
10 & -6.6 & -1.89\cdot 10^{-3} & -0.11 & 0.7 & -0.18 & 122 \\
11 & -3.3 & -9.6\cdot 10^{-4} & -6.6\cdot 10^{-2} & 12.8 & -0.09 & 122 \\
12 & 8.1 & 2.8\cdot 10^{-3} & 0.44 & 26.6 & 0.83 & 128 \\
13 & -7.7 & -2.6\cdot 10^{-3} & -0.12 & -4.95 & -0.81 & 128\\
14 & 13.2 & 4.3 \cdot 10^{-3} & 1.6 & 510 & 1.23 & 128 \\
15 & -3.8 & -1.2\cdot 10^{-4} & 0.7 & 153 & -0.36 & 128\\
16 & 4.3 & 1.13 \cdot10^{-3} & 1.4\cdot 10^{-2} & -1.35 & 0.46 & 134 \\
17 & -2.09 & -5.53\cdot10^{-4} & -8.24\cdot 10^{-2} & 9.43 & -0.23 & 134 \\
18 & 6.43 & 1.67\cdot 10^{-3} & 1.01 & 259 & 0.67 & 134 \\
19 & -0.49 & -1.27\cdot 10^{-4} & 0.76 & 164 & -5.7\cdot 10^{-2} & 134 \\
\hline
\end{array}
$$
\end{footnotesize}

\bigskip
$$
{\rm Table\ 1\ (b)}
$$
\newpage

\vskip 2in

\begin{center}
\setlength{\unitlength}{0.6mm}
\begin{picture}(140,140)
\put(0,0){\framebox(154,154){$\mbox{ }$}}
\multiput(2,0)(30,0){6}{\line(0,1){1}}
\multiput(2,154)(30,0){6}{\line(0,-1){1}}
\multiput(154,2)(0,25){7}{\line(1,0){1}}
\multiput(0,2)(0,25){7}{\line(1,0){1}}
\put(-2,2){\makebox(0,0)[r]{110}}
\put(-2,27){\makebox(0,0)[r]{115}}
\put(-2,52){\makebox(0,0)[r]{120}}
\put(-2,77){\makebox(0,0)[r]{125}}
\put(-2,102){\makebox(0,0)[r]{130}}
\put(-2,127){\makebox(0,0)[r]{135}}
\put(-2,152){\makebox(0,0)[r]{140}}
\put(2,-2){\makebox(0,0)[t]{160}}
\put(32,-2){\makebox(0,0)[t]{165}}
\put(62,-2){\makebox(0,0)[t]{170}}
\put(92,-2){\makebox(0,0)[t]{175}}
\put(122,-2){\makebox(0,0)[t]{180}}
\put(152,-2){\makebox(0,0)[t]{185}}
\put(77,-12.5){\makebox(0,0){$m_t(m_t)$ [GeV]}}
\put(77,-35){\makebox(0,0){\footnotesize Fig. 1}}
\put(-15,77){\makebox(0,0)[r]{$M_{h^0}$}}
\put(-15,67){\makebox(0,0)[r]{[GeV]}}
\thicklines
\put(2,31){\line(3,2){150}}
\end{picture}
\end{center}

\end{document}